\documentclass[pra,twocolumn,superscriptaddress,10pt,showpacs]{revtex4-1}
\usepackage{graphicx,epstopdf}
\usepackage{amsmath}
\usepackage{hyperref}
\usepackage{amsfonts}
\usepackage{bbm}
\usepackage{amssymb}
\usepackage{color}
\usepackage{latexsym}
\usepackage{times,txfonts}
\newtheorem{theorem}{Theorem}

\usepackage[usenames,dvipsnames]{xcolor}

\newcommand{\1}{\mathbbm{1}}
\definecolor{DarkGreen}{RGB}{000,70,000}

\begin{document}

\title{Shortcut to adiabatic gate teleportation}

\author{Alan C. Santos}
\email{alancs@if.uff.br}
\affiliation{Instituto de F\'{i}sica, Universidade Federal Fluminense, Av. Gal. Milton Tavares de Souza s/n, Gragoat\'{a}, 24210-346 Niter\'{o}i, Rio de Janeiro, Brazil}
\author{Raphael D. Silva}
\email{raphael.dias.1987@gmail.com}
\affiliation{Instituto de F\'{i}sica, Universidade Federal Fluminense, Av. Gal. Milton Tavares de Souza s/n, Gragoat\'{a}, 24210-346 Niter\'{o}i, Rio de Janeiro, Brazil}
\author{Marcelo S. Sarandy}
\email{msarandy@if.uff.br}
\affiliation{Instituto de F\'{i}sica, Universidade Federal Fluminense, Av. Gal. Milton Tavares de Souza s/n, Gragoat\'{a}, 24210-346 Niter\'{o}i, Rio de Janeiro, Brazil}
\affiliation{Center for Quantum Information Science \& Technology and Ming Hsieh Department of Electrical Engineering, University of Southern California, Los Angeles, California 90089, USA}

\begin{abstract}
We introduce a shortcut to the adiabatic gate teleportation model of quantum computation. 
More specifically, we determine fast local counter-diabatic Hamiltonians able to implement  
teleportation as a universal computational primitive. In this scenario, we provide 
the counter-diabatic driving for arbitrary $n$-qubit gates, which allows to achieve 
universality through a variety of gate sets. Remarkably, our approach maps the superadiabatic 
Hamiltonian $H_{\text{SA}}$ for an arbitrary $n$-qubit {\it gate} teleportation into the implementation of 
a rotated superadiabatic dynamics of an $n$-qubit {\it state} teleportation. 
This result is rather general, with the speed of the evolution only dictated by the quantum 
speed limit. In particular, we analyze the energetic cost for different Hamiltonian interpolations in the 
context of the energy-time complementarity. 
\end{abstract}


\maketitle

\section{Introduction}

Quantum teleportation~\cite{Bennett} is a valuable tool for a number of quantum tasks. In quantum communication, 
it makes available a quantum channel for transmission of unknown states between two agents (Alice and Bob) 
separated by a large distance (currently more than $100$ km in optical fibers~\cite{Takesue:15} or $143$ km in a free-space link~\cite{Ma:12}). 
In quantum information processing, 
quantum teleportation can be applied as a primitive for universal quantum computation (QC), as remarkably shown 
by Gottesman and Chuang in Ref.~\cite{Gottesman-Chuang}. In this approach, a third party (Charlie) provides rotated Bell states to Alice and Bob, 
who can implement universal QC by solely performing single-qubit operations and Bell measurements. In particular, this method 
is a precursor of the paradigm of measurement-based QC (see, e.g., Ref.~\cite{Briegel:09}). More recently, QC via quantum teleportation has been formulated via 
adiabatic evolution by Bacon and Flammia~\cite{Bacon-Flammia}, providing a hybrid approach for QC (see also Ref.~\cite{Hen:15} for an alternative adiabatic hybrid approach). 
In this scenario, a quantum circuit can be mapped in a sequence 
of piecewise Hamiltonian evolutions implementing single- and double-gate teleportation protocols, allowing for universality  
through the set of one-qubit rotations joint with an entangling two-qubit gate~\cite{Barenco:95,Nielsen:book} . However, since these processes 
are ruled by the adiabatic approximation, it turns out that each gate of the adiabatic circuit will be implemented within some fixed probability (for a finite evolution time). Moreover, the time for performing each individual gate will be bounded from below by the adiabatic time condition~\cite{Messiah:book} . 

In order to speed up the adiabatic evolution in the Bacon-Flammia hybrid model, we propose here a general shortcut to 
adiabatic gate teleportation via counter-diabatic assistant Hamiltonians within the framework of the superadiabatic 
theory~\cite{Demirplak:03,Demirplak:05,Berry:09,Torrontegui:13}. 
In particular, we introduce the concept of superadiabatic gate teleportation, showing that it can be used as a fast primitive 
for universal QC. The use of superadiabatic evolutions for universal QC via local interactions 
has recently been proposed in Ref.~\cite{alan-sarandy}, where it is shown how to implement arbitrary $n$-controlled gates with minima ancilla requirements. 
The physical resources spent by this strategy will be governed by the  
quantum circuit complexity, but no adiabatic constraint will be required in the individual implementation of the quantum gates. 
Moreover, the gates will be deterministically implemented with probability one as long as decoherence effects can be avoided. 
This analog approach allows for fast implementation of individual gates, whose time 
consumption is only dictated by the quantum speed limit (QSL) (for closed systems, see Refs.~\cite{Mandelstam:45,Margolus:98,Giovannetti:03,Deffner:13}). 
Indeed, the time demanded for each gate will imply an energy cost, which increases with the desired speed of the evolution. 

The paper is organized as follows. In Sec. \ref{secaoII}, we discuss the adiabatic gate teleportation protocol as originally 
proposed in Ref.~\cite{Bacon-Flammia}, by explictly extending it to arbitrary $n$-qubit gates. In Sec. \ref{secaoIII}, we 
derive a shortcut for the adiabatic teleportation of $n$-qubit gates, showing that it can be used to implement universal QC. 
Moreover, since no adiabaticity is required, we also analyze the energetic cost for implementing superadiabatic universal QC via 
adiabatic gate teleportation. Section~\ref{Conc} is devoted to our conclusions. 

\section{Universal QC via adiabatic teleportation} \label{secaoII}

\subsection{Adiabatic teleportation of one-qubit states} \label{secao1.1}
\begin{figure}[!htb]
\centering
\includegraphics[scale=0.44]{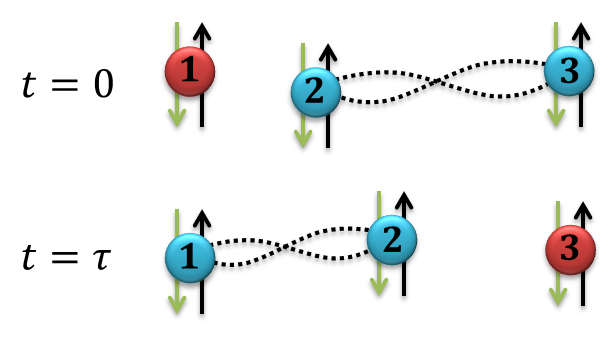}
\caption{(Color online) Adiabatic teleportation of a single qubit (red particle). The quantum state initially encoded in qubit 1 ($t=0$) is teleported 
to qubit 3 ($t=\tau$), with a Bell pair (blue particles)  used a resource for the protocol.}
\label{figure1}
\end{figure}

Given an unknown state $\left\vert \psi \right\rangle =a\left\vert
0\right\rangle +b\left\vert 1\right\rangle $, where $\left\vert a\right\vert
^{2}+\left\vert b\right\vert ^{2}=1$, adiabatic teleportation can be implemented through the Hamiltonian~\cite{Bacon-Flammia}
\begin{equation}
H_{0}\left( s\right) =\eta _{i}\left( s\right) H_{i}+\eta _{f}\left(
s\right) H_{f}  \, , \label{BF.1.1}
\end{equation}%
where $\eta _{i}\left( 0\right) =\eta _{f}\left( 1\right) =1$, $\eta
_{i}\left( 1\right) =\eta _{f}\left( 0\right) =0$, and
\begin{eqnarray}
H_{i} = -\omega \hbar \left( \mathbbm{1}XX+\mathbbm{1}ZZ\right) \, ,
\label{BF.1.2a} \\
H_{f} = -\omega \hbar \left( XX\mathbbm{1}+ZZ\mathbbm{1}\right) \, ,
\label{BF.1.2b}
\end{eqnarray}
where $X$ and $Z$ are Pauli spin-$\frac{1}{2}$ operators and $s=t/\tau$ is the normalized time, with $\tau$ the total evolution time. The state of the system at $t=0$ is prepared as $\left\vert \phi (0) \right\rangle =(1/\sqrt{2})\left\vert \psi \right\rangle \left(\left\vert 00 \right\rangle+\left\vert 11 \right\rangle\right)$. To prove that teleportation happens, we must show 
that the final state of the system is given by $(1/\sqrt{2})\left(\left\vert 00 \right\rangle+\left\vert 11 \right\rangle\right)\left\vert \psi \right\rangle$.  A scheme of the process is shown in Fig.~\ref{figure1}.

It is important to notice that the Hamiltonian $H_{0}\left( s\right)$ acts only on qubits 2 and 3 for $s=0$, only on qubits 1 and 2 for $s=1$,  
and on all the 3 qubits for $0<s<1$. Since $H_{0}\left( s\right)$ is doubly degenerate, the adiabatic theorem implies solely in the 
decoupled evolution of the eigenspaces of $H_{0}\left( s\right)$. Then, in order to show the success of the adiabatic teleportation via 
$H_{0}\left( s\right)$, Bacon and Flammia~\cite{Bacon-Flammia} proceeded by developing an analysis based on logical qubits. Here, we 
devise an alternative derivation, which is based directly on the symmetries of $H_0(s)$. First, consider the  commutation relations
\begin{equation}
\left[ H_0\left( s\right) ,\Pi _{z}\right] =\left[ H_0\left( s\right) ,\Pi _{x}%
\right] =0 \, , \label{BF.1.3a}
\end{equation}%
with $\Pi _{z}=ZZZ$ and $\Pi _{x}=XXX$. For a state of the computational basis $\{\left\vert nmk\right\rangle\}$ we have%
\begin{eqnarray}
\Pi _{z}\left\vert nmk\right\rangle &=&\left( -1\right) ^{n+m+k}\left\vert
nmk\right\rangle \, , \label{BF.1.4a} \\
\Pi _{x}\left\vert nmk\right\rangle &=&\left\vert \bar{n}\bar{m}\bar{k}%
\right\rangle , \label{BF.1.4b}
\end{eqnarray}
where we have defined $|\bar{0}\rangle \equiv |1\rangle$ and $|\bar{1}\rangle \equiv |0\rangle$.  
Notice that $\Pi _{z}$ and $\Pi _{x}$ are parity operators, each of them associated with a $Z_2$ symmetry of the Hamiltonian. 
Now, let us define the sets $\left\{ | nmk\rangle_\pm \right\} $ given by vectors of the computational basis with $\Pi _{z}$ eigenvalues $\pm 1$. 
Then, from the commutation of the Hamiltonian $H_{0}\left( s\right)$  
with $\Pi _{z}$, we obtain that parity is conserved throughout the evolution, which means that we can conveniently write $H_{0}\left( s\right)$ in 
a block-diagonal basis
\begin{equation}
H_{0}\left( s\right) =\left( 
\begin{array}{cc}
H_{4\times 4}^{+}\left( s\right)  & \emptyset _{4\times 4} \\ 
\emptyset _{4\times 4} & H_{4\times 4}^{-}\left( s\right) 
\end{array}%
\right) ,  \label{BF.1.5}
\end{equation}%
where the basis has been ordered in terms of $\left\{ | nmk\rangle_{+} , | nmk\rangle_{-} \right\} $. 
In addition, the symmetry $\Pi _{x}$ ensures a relationship between the elements of $H_{4\times 4}^{+}\left(
s\right) $ and $H_{4\times 4}^{-}\left( s\right) $ so that, if we conveniently sort the computational basis in the parity subspaces 
$\left\{ | nmk\rangle_{+} \right\}$ and 
$\left\{ | nmk\rangle_{-} \right\} $,  we find that $H_{4\times 4}^{+}\left( s\right) =H_{4\times
4}^{-}\left( s\right) $. In fact, by computing the matrix elements of $H_{4\times 4}^{+}\left( s\right) $ and $H_{4\times 4}^{-}\left( s\right) $
and by using that $\Pi _{x} | nmk\rangle_+ = | \bar{n}\bar{m}\bar{k}\rangle_{-}$, we get 
\begin{equation}
_{-}\langle \bar{n}^{\prime }\bar{m}^{\prime }\bar{k}^{\prime} | H_{0}\left( s\right) |\bar{n}\bar{m}\bar{k}\rangle_{-} \, = \, 
_{+}\langle n^{\prime }m^{\prime }k^{\prime } | H_{0} \left( s\right)  |nmk\rangle_{+}  \,\, .
\label{BF.1.7}
\end{equation}%

Then, by computing the spectrum of $H_{4\times
4}^{\pm }\left( s\right) $, we completely  determine the spectrum of $H_{0}\left(
s\right)$. More specifically, the energies associated with $H_{4\times
4}^{\pm }\left( s\right) $ read as
\begin{eqnarray}
E_{0}\left( s\right)  &=&-2\omega \hbar \sqrt{\eta _{i}^{2}\left( s\right)
+\eta _{f}^{2}\left( s\right) } \, , \label{BF.1.9a} \\
E_{1}\left( s\right)  &=&E_{2}\left( s\right) =0  \label{BF.1.9b}   \, , \\
E_{3}\left( s\right)  &=&2\omega \hbar \sqrt{\eta _{i}^{2}\left( s\right)
+\eta _{f}^{2}\left( s\right) } , \label{BF.1.9c}
\end{eqnarray}%
with the gap between the ground state and the first excited state given by
\begin{equation}
\varepsilon \left( s\right) =2\omega \hbar \sqrt{\eta _{i}^{2}\left(
s\right) +\eta _{f}^{2}\left( s\right) } . \label{BF.1.10}
\end{equation}

We can observe that $\varepsilon\left( s\right) \neq 0$ $\forall s\in \left[ 0,1\right] $
because $\eta _{i}\left( s\right) $ and $\eta _{f}\left( s\right) $ never
simultaneously vanish. To conclude the teleportation of the initial state, it remains to show that the final state of the third qubit is exactly 
$\left\vert \psi \right\rangle $. To this end, let us write the initial and final states as%
\begin{eqnarray}
\left\vert \phi \left( 0\right) \right\rangle &=&\frac{1}{\sqrt{2}}\left(
a\left\vert 0\right\rangle _{1}+b\left\vert 1\right\rangle _{1}\right)
\left( \left\vert 00\right\rangle _{23}+\left\vert 11\right\rangle
_{23}\right)  \label{BF.1.11a} , \\
\left\vert \phi \left( 1\right) \right\rangle &=&\frac{1}{\sqrt{2}}\left(
\left\vert 00\right\rangle _{12}+\left\vert 11\right\rangle _{12}\right)
\left( \alpha \left\vert 0\right\rangle _{3}+\beta \left\vert 1\right\rangle
_{3}\right) , \label{BF.1.11b}
\end{eqnarray}%
where the form of $\left\vert \phi \left( 1\right) \right\rangle$ is ensured by the adiabatic theorem, with general coefficients $\alpha =\alpha \left( a,b\right) $ and $\beta =\beta
\left( a,b\right) $. Now notice that Eq.~(\ref{BF.1.11a}) implies that the coefficients $a$ and $b$ 
multiply the states of parity $+1 $ and $-1$, respectively. In addition, Eq.~(\ref{BF.1.11b}) implies that 
the coefficients $\alpha \left( a,b\right) $ and $\beta \left( a,b\right) $
also multiply states of parity $+1$ and $-1$, respectively. Due to the symmetry 
$\Pi _{z}$, it follows that states of different parities evolve independently. 
Then, $%
\alpha =\alpha \left( a\right) $ and $\beta =\beta \left( b\right) $. Moreover, since 
the evolution of the system is unitary, we have that $%
\left\langle \phi \left( 0\right) |\phi \left( 0\right) \right\rangle
=\left\langle \phi \left( 1\right) |\phi \left( 1\right) \right\rangle =1$. This implies that  $\left\vert \alpha \left( a\right)
\right\vert ^{2}=\left\vert a\right\vert ^{2}$ and $\left\vert \beta \left(
b\right) \right\vert ^{2}=\left\vert b\right\vert ^{2}$. Consequently, $%
\alpha \left( a\right) =ae^{i\theta _{a}}$ and $\beta \left( a\right)
=be^{i\theta _{b}}$, for any $\theta _{a}$ and $\theta _{b}$ real. 
On the other hand, we can use the parity $\Pi _{x}$ to show that states of
parities $+1$ and $-1$ have identical evolution, since $H_{4\times 4}^{+}\left( s\right) =H_{4\times
4}^{-}\left( s\right) $. Then, $\theta _{a}=\theta _{b}=\theta$. Hence,
\begin{equation}
\left\vert \phi \left( 1\right) \right\rangle =\frac{1}{\sqrt{%
2}}\left( \left\vert 00\right\rangle _{12}+\left\vert 11\right\rangle
_{12}\right) \left( a\left\vert 0\right\rangle _{3}+b\left\vert
1\right\rangle _{3}\right) \, , \label{estado-final-simples}
\end{equation}%
up to a global phase $e^{i\theta}$. This concludes the proof of the adiabatic teleportation 
of a single qubit. 

\subsection{Adiabatic teleportation of  $n$-qubit states}  \label{AGT-N}

\begin{figure}[!htb]
\centering
\includegraphics[scale=0.44]{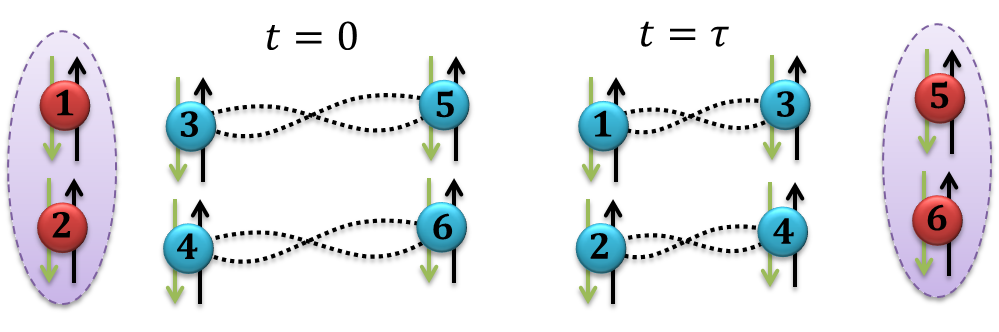}
\caption{(Color online) Adiabatic teleportation of a two-qubit state (red particles) .  
The composite state to be teleported is prepared in qubits 1 and 2, with the final state of teleportation in qubits 5 and 6. Two Bell pairs (blue particles) 
are used in the protocol.}
\label{figure2}
\end{figure}

Let us begin by generalizing the previous protocol to implement now the adiabatic teleportation of an unknown two-qubit state. In this direction,  
we will consider a quantum system composed of six qubits. 
A scheme of the process is exhibited in Fig.~\ref{figure2}. The composite state to be teleported is prepared in qubits $1$ and $2$ and 
the final state in qubits 5 and 6, with two Bell pairs used as the resource for the protocol.
Let us write the state to be teleported as%
\begin{equation}
\left\vert \psi \right\rangle _{12}=\alpha \left\vert 00\right\rangle
_{12}+\delta \left\vert 01\right\rangle _{12}+\gamma \left\vert
10\right\rangle _{12}+\beta \left\vert 11\right\rangle _{12} \, , \label{BF.2.0}
\end{equation}

The adiabatic teleportation of the initial state will be performed through the Hamiltonian 
\begin{equation}
H_{D}\left( s\right) =\mathbbm{1}_{\text{even}} \otimes H_{\text{odd}}\left( s\right)
+H_{\text{even}}\left( s\right) \otimes \mathbbm{1}_{\text{odd}} \, , \label{BF.2.1}
\end{equation}%
where $H_{\text{even}}\left( s\right)$ and $H_{\text{odd}}\left( s\right)$ are given by $H_0(s)$ as given by Eq.~(\ref{BF.1.1}) acting over qubits 
labeled with {\it even} and {\it odd} indices, respectively. 
Then, no interaction between the odd and even sectors will occur. To determine the spectrum of $H_{D}\left(
s\right) $ we will make use of the following general result:
Let us consider $A_{m \times m}$ and $B_{n \times n}$ as two operators such that
$A_{m\times m}\left\vert a_{\mu }\right\rangle
=a_{\mu }\left\vert a_{\mu }\right\rangle $ and $B_{n\times n}\left\vert
b_{\eta }\right\rangle =b_{\eta }\left\vert b_{\eta }\right\rangle $, with the sets of eigenvalues $\{b_{\eta }\}$ and $\{a_{\mu }\}$ associated with the eigenvector 
bases $\{\left\vert
b_{\eta }\right\rangle\}$ and $\{\left\vert a_{\mu }\right\rangle\}$, respectively. Thus, if we consider an operator $C_{k\times k}$, where 
$k=mn$, such that $C_{k\times k}=A_{m\times m}\otimes \mathbbm{1}_{n\times n}+\mathbbm{1}_{m\times
m} \otimes B_{n\times n}$, then $\left\vert c_{\mu \eta }\right\rangle =\left\vert
a_{\mu }\right\rangle \otimes  \left\vert b_{\eta }\right\rangle $ are the
eigenvectors of $C_{k\times k}$ associated with the eigenvalues $c_{\mu \eta
}=a_{\mu }+b_{\eta }$.
Bearing in mind this result, the spectrum of $H_{D}\left( s\right) $ is simply 
given by%
\begin{equation}
E_{kl}\left( s\right) =E_{k}^{\text{odd}}\left( s\right) +E_{l}^{\text{even}}\left(
s\right) \, , \label{BF.2.2}
\end{equation}%
where $E_{k}^{\text{odd}}\left( s\right) $ and $E_{k}^{\text{even}}\left( s\right) $ are
given by Eqs.~(\ref{BF.1.9a}),~(\ref{BF.1.9b}), and~(\ref{BF.1.9c}). By
using Eq.~(\ref{BF.2.2}), we show that the gap of
the $H_{D}\left( s\right) $ is $\varepsilon _{D}\left( s\right)
=E_{01}\left( s\right) -E_{00}\left( s\right) =\varepsilon \left( s\right) $%
, where $\varepsilon \left( s\right) $ was determined by Eq.~(\ref{BF.1.10}). 
As each sector has the symmetries $\Pi _{x}$ and $\Pi _{z}$, we define the operators%
\begin{eqnarray*}
\Pi _{z\text{\ }\text{odd}} &\equiv &\mathbbm{1} \otimes \Pi _{z} \text{ \ \ , \ \ }
\Pi_{z\text{\ }\text{even}} \equiv \Pi _{z} \otimes \mathbbm{1} , \\
\Pi _{x\text{\ }\text{odd}} &\equiv &\mathbbm{1} \otimes \Pi _{x} \text{ \ \ , \ \ }
\Pi_{x\text{\ }\text{even}}\equiv \Pi _{x} \otimes \mathbbm{1} ,
\end{eqnarray*}%
where the left operators in the tensor product act on the even sector, with the right operators acting on the odd sector. 
It then follows that these operators (and their) products are $Z_2$ symmetries of $H_{D}\left( s\right) $. 
Considering the symmetry operator $\Pi _{z}^{D}=\Pi _{z\text{\ }\text{even}}\,\Pi _{z\text{\ }\text{odd}}$, we then write%
\begin{equation}
H_{D}\left( s\right) =\left( 
\begin{array}{cc}
H_{32\times 32}^{+}\left( s\right) & \emptyset _{32\times 32} \\ 
\emptyset _{32\times 32} & H_{32\times 32}^{-}\left( s\right)%
\end{array}%
\right) \, ,
\end{equation}%
where $H_{32\times 32}^{\pm }\left( s\right) $ acts on the states of parity $\pm 1$ of the operator $\Pi _{z}^{D}$. 
By using now the symmetry $\Pi _{x}^{D}=\Pi _{x\text{\ }\text{even}}\Pi _{x\text{\ }\text{odd}}$ , we can choose the order of the basis such that 
$H_{32\times 32}^{+}\left( s\right)=H_{32\times 32}^{-}\left( s\right) $. In addition, by using the
symmetries $\Pi _{z\text{\ }\text{odd}}$ and $\Pi _{z\text{\ }\text{even}}$ of each sector we get
\begin{equation}
H_{D}\left( s\right) =\left( 
\begin{array}{cccc}
H_{\alpha }\left( s\right) & \emptyset & \emptyset & \emptyset \\ 
\emptyset & H_{\beta }\left( s\right) & \emptyset & \emptyset \\ 
\emptyset & \emptyset & H_{\gamma }\left( s\right) & \emptyset \\ 
\emptyset & \emptyset & \emptyset & H_{\delta }\left( s\right)%
\end{array}%
\right) , \label{-.2.9}
\end{equation}%
where we have considered the specific parity ordering 
$\left\{ \left\vert E\right\rangle_{+} \left\vert O\right\rangle_{+}  , 
\left\vert E\right\rangle_{-} \left\vert O\right\rangle_{-}  ,
\left\vert E\right\rangle_{-} \left\vert O\right\rangle_{+} , 
\left\vert E\right\rangle_{+} \left\vert O\right\rangle_{-}    
\right\} $ in the computational basis, 
with the definitions 
$ \left\vert E \right\rangle  \equiv \left\vert n_{2} n_{4} n_{6} \right\rangle$ and
$ \left\vert O \right\rangle \equiv \left\vert n_{1} n_{3} n_{5} \right\rangle$. 
Moreover, by using the symmetries of $H_D(s)$ with respect to $\Pi _{x\text{\ }\text{odd}}$ and $\Pi _{x\text{\ }\text{even}}$, 
we find that the blocks $\left\{ H_\alpha(s), H_\beta(s), H_\gamma(s), H_\delta(s) \right\}$ are identical by a suitable organization 
of the basis vectors. 

\begin{figure}[!htb]
\centering
\includegraphics[scale=0.36]{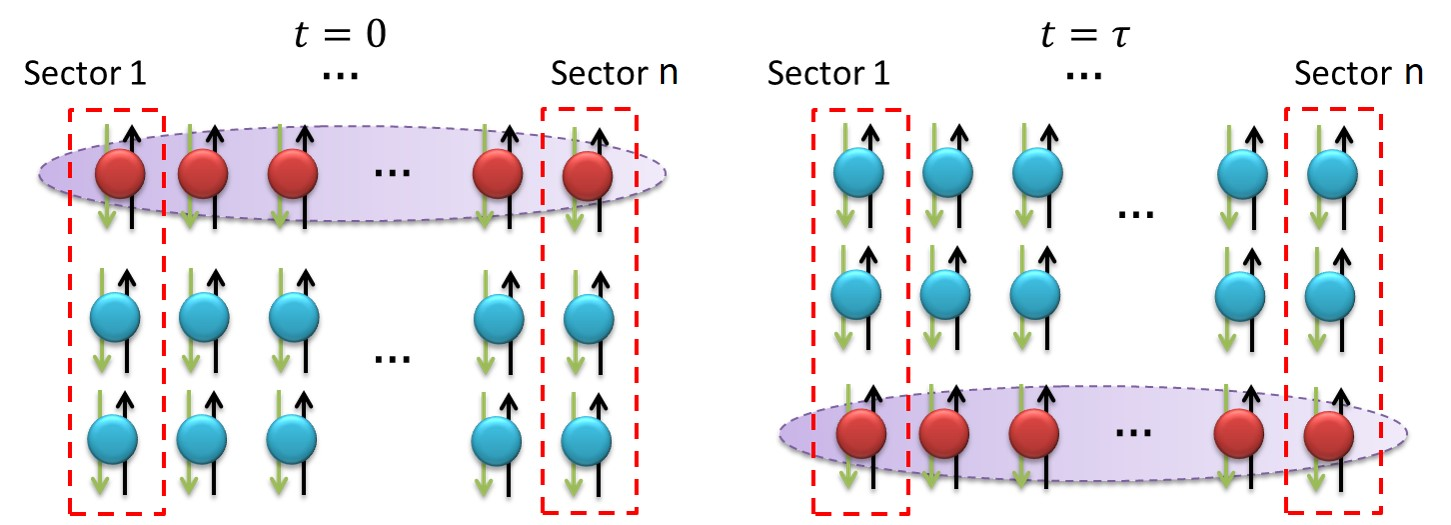}
\caption{(Color online) Adiabatic teleportation of an $n$-qubit state. 
In each three-qubit sector we have a qubit to be teleported (red particle) and a Bell pair (blue particles).}
\label{figure3}
\end{figure}

To show that double
teleportation can indeed be adiabatically implemented via the Hamiltonian $H_{D}\left( s\right) $, let us 
denote the initial and final states as given by%
\begin{eqnarray}
\left\vert \phi \left( 0\right) \right\rangle &=&\left\vert \psi
\right\rangle _{12}\left\vert \beta _{00}\right\rangle _{35}\left\vert \beta
_{00}\right\rangle _{46}  , \label{BF.2.11.a} \\
\left\vert \phi \left( 1\right) \right\rangle &=&\left\vert \beta
_{00}\right\rangle _{12}\left\vert \beta _{00}\right\rangle _{13}\left\vert 
\tilde{\psi}\right\rangle _{56} , \label{BF.2.11.b}
\end{eqnarray}
where $\left\vert \beta _{00} \right\rangle = 1/\sqrt{2} \left( \left\vert 00 \right\rangle + \left\vert 11 \right\rangle \right) $ and 
 $\left\vert \tilde{\psi}\right\rangle _{56}$ reads as
\begin{equation}
\left\vert \psi \right\rangle _{56}=\tilde{\alpha}\left\vert 00\right\rangle
_{56}+\tilde{\delta}\left\vert 01\right\rangle _{56}+\tilde{\gamma}%
\left\vert 10\right\rangle _{56}+\tilde{\beta}\left\vert 11\right\rangle
_{56}  . \label{BF.2.12}
\end{equation}
Note that, since $H_D(s)$ is degenerate, we cannot associate $| \psi \rangle_{56}$ directly to $| \psi \rangle_{12}$. 
However, Eq.~(\ref{-.2.9}) implies into a dynamics such as $\tilde{\xi}=\tilde{\xi}%
\left( \xi \right) $, where $\tilde{\xi}=\left\{ \tilde{\alpha},\tilde{\delta%
},\tilde{\gamma},\tilde{\beta}\right\} $ and $\xi =\left\{ \alpha ,\delta
,\gamma ,\beta \right\} $. This is because each element of the set $\left\{ \alpha ,\delta
,\gamma ,\beta \right\}$ is in a distinct parity sector. Moreover, unitarity of the evolution leads to $\left\vert \tilde{\xi}\left( \xi \right)
\right\vert ^{2}=$ $\left\vert \xi \right\vert ^{2}$, which yields 
 $\tilde{\xi}=$ $\xi e^{i\varphi _{\xi }}$. 
By using now the parity operators $\Pi _{x\text{\ }\text{odd}}$ and $\Pi _{x\text{\ }\text{even}}$, we can show that the blocks in the Hamiltonian 
provided by Eq.~(\ref{-.2.9}) are identical (by suitably ordering the basis) so that the parameters $\varphi
_{\xi }$ are globally defined, namely, $\varphi_{\xi } \equiv \varphi$ $(\forall \text{\ } \xi$). Hence, we
conclude that the state of the qubits $5$ and $6$ at the final of the process reads as
\begin{equation}
\left\vert \psi \right\rangle _{56}= \alpha
\left\vert 00\right\rangle _{56}+\delta \left\vert 01\right\rangle
_{56}+\gamma \left\vert 10\right\rangle _{56}+\beta \left\vert
11\right\rangle _{56}  \, , \label{estado-final-duplo}
\end{equation}%
up to the global phase $e^{i\varphi}$. We can extend this protocol to perform teleportation of an unknown state of $n$ qubits. 
In this direction, we need to increase the number of sectors and define a Hamiltonian given by 
$H_{\text{mult}}\left( t\right)=\sum_{k=1}^{n}H_{k}\left( t\right)$, where each $H_{k}\left( t\right)$ is given by Eq.~(\ref{BF.1.1}), which 
acts on an individual sector composed by three qubits. Consequently, $n$ Bell pairs will be used as a resource for the process. 
A scheme of such generalized protocol is presented in Fig.~\ref{figure3}. 
The Hamiltonian $H_{\text{mult}}(t)$ displays a $2n$-fold degenerate ground state, which decouples from the rest of the spectrum in the adiabatic dynamics. 
Teleportation of the $n$-qubit state will then follow from the $z$ and $x$ parity symmetries in each individual sector.

\subsection{Adiabatic teleportation of unitary $n$-qubit gates}
In the one-qubit gate teleportation protocol, Alice starts with an unknown state $\left\vert \psi \right\rangle$ at qubit 1 and shares a 
rotated Bell pair $U_{3}\left\vert \beta _{00}\right\rangle _{23}$ with Bob (prepared by a third party Charlie). Then, by applying the 
usual teleportation procedure, Bob receives $U_3\left\vert \psi \right\rangle$ at the end of the protocol with probability one as long as 
decoherence can be neglected. In order to implement the adiabatic version of gate teleportation, 
we define the gate to be implemented over qubit 3 as  
$U=\mathbbm{1}_{1}\mathbbm{1}_{2}U_{3}$, where $U_{3}^{\dag }U_{3}=\mathbbm{1}_{3}$. 
Then, as shown in Ref.~\cite{Bacon-Flammia}, the time-dependent Hamiltonian $H_{0}\left( s,U\right)$ able to adiabatically implement the teleportation 
of the gate $U$ can be determined from the original Hamiltonian $H_0(s)$ for one-qubit teleportation through the rotation 
\begin{equation}
H_{0}\left( s,U\right) =UH_{0}\left( s\right) U^{\dag }  .  \label{a.1.1}
\end{equation}%
Indeed, this can be understood directly from the symmetries of $H_{0}\left( s,U\right)$. 
Since commutation relations are preserved by rotations~\cite{Sakurai}, $H_{0}\left( s,U\right)$ 
is $Z_2$-symmetric under the parity operators $\Pi_{z}\left( U\right) =ZZ\left( U_{3}ZU_{3}^{\dag }\right)$ 
and $\Pi_{x}\left( U\right) =XX\left( U_{3}XU_{3}^{\dag }\right) $. Then, we can show the 
teleportation of the gate $U$ by working the computational basis rotated by $U$. In this new basis, 
the matrix form of $H_{0}\left( s,U\right)$ is identical to that of the original $H_{0}\left( s\right)$, which 
implies that the same argument used to the simple teleportation performed by $H_{0}\left(
s\right) $ is applicable to case of the Hamiltonian $H_{0}\left( s,U\right) $. 
The gap of $H_{0}\left( s,U\right)$ is also given by $\left( \ref{BF.1.10}%
\right) $ because the spectrum of the operator will not change by a unitary
transformation~\cite{Sakurai}. Hence, the initial state 
$\left\vert \phi \left( 0,U\right) \right\rangle =\left\vert
\psi \right\rangle _{1}U_{3}\left\vert \beta _{00}\right\rangle _{23}$ (with the rotated Bell pair provided by Charlie) 
will be adiabatically evolved into the final state 
$\left\vert\phi \left( 1,U\right) \right\rangle =\left\vert \beta _{00}\right\rangle
_{12}U_{3}\left\vert \psi \right\rangle _{3}$.

In order to perform universal QC via adiabatic teleportation, Ref.~\cite{Bacon-Flammia} specifically worked out 
a Hamiltonian to adiabatically implement the teleportation of the controlled-phase gate. Here, we extend the 
protocol to adiabatically implement an arbitrary $n$-qubit unitary gate. By focusing first on two-qubit gates, 
we use as a  fundamental resource the double teleportation protocol, as described in Sec. \ref{AGT-N}. 
More specifically, we can show that any two-qubit gate $U$ can be implemented by the Hamiltonian%
\begin{equation*}
H_{D}\left( s,U\right) =UH_{D}\left( s\right) U^{\dag } , 
\end{equation*}%
where $H_{D}\left( s\right) $ is provided by Eq.~(\ref{BF.2.1}) and $U=U_{56}$ is the gate to be performed 
at the final time in the qubits of the Bob. As in the case of single qubits, we have that the spectra of $%
H_{D}\left( s,U\right) $ and $H_{D}\left( s\right) $ are identical. Then, to show that the two-qubit gate teleportation 
takes place through the adiabatic dynamics dictated by $H_{D}\left( s,U\right)$, we make use of the following 
rotated parity symmetry operators:
\begin{eqnarray*}
\Pi _{z\text{\ }\text{sec}}\left( U\right) &=&U\Pi _{z\text{\ }\text{sec}}^{D}U^{\dag }\text{ \ \ , \ \ }\Pi
_{x\text{\ }\text{sec}}\left( U\right) =U\Pi _{z\text{\ }\text{sec}}^{D}U^{\dag } , \\
\Pi _{x}\left( U\right) &=&U\Pi _{x}^{D}U^{\dag }\text{ \ \ \ \ \ , \ \ \ \
\ }\Pi _{z}\left( U\right) =U\Pi _{z}^{D}U^{\dag } ,
\end{eqnarray*}%
where $\text{sec}=\left\{ \text{even},\text{odd}\right\} $. Bearing in mind that Charlie provides rotated Bell pairs, we have at 
$s=0$ the initial state $\left\vert \phi \left(
0,U\right) \right\rangle =U_{56}\left\vert \phi \left( 0\right)
\right\rangle $, where $\left\vert \phi \left( 0\right) \right\rangle $ is
given by Eq.~(\ref{BF.2.11.a}). In the  rotated basis, the matrix 
form of $H_{D}\left( s,U\right)$ is also equivalent to the matrix form of $%
H_{D}\left( s\right)$ as given in the original basis, from which it follows that at the final of
the process the state of the system will by $\left\vert \phi \left(
1,U\right) \right\rangle =U_{56}\left\vert \phi \left( 1\right)
\right\rangle $, where $\left\vert \phi \left( 1\right) \right\rangle $ is given by Eq.~(\ref{estado-final-duplo}).
Concerning the adiabatic teleportation of an $n$-qubit gate $U_n$, it can be implemented from the simple 
adiabatic teleportation of an $n$-qubit state, as previously described. The Hamiltonian that adiabatically 
implements this task is then $H_{\text{mult}}\left( t,U_{n}\right)=U_{n} H_{\text{mult}}\left( t\right) U_{n}^{\dag }$. 
This allows for universal QC by using a variety of sets of universal gates, e.g., the set composed by Hadamard 
added by three-qubit Toffoli gates~\cite{Shi:03,Aharonov:03}.

\section{Superadiabatic QC via teleportation} \label{secaoIII}

\subsection{Shortcut to adiabaticity} 

We can obtain fast piecewise implementation of quantum gates via shortcuts to 
adiabaticity~\cite{Demirplak:03,Demirplak:05,Berry:09,Torrontegui:13}, whose evolution 
time will not be constrained by the adiabatic theorem. We begin by defining the evolution operator
\begin{equation}
U\left( t\right) =\sum_{n}e^{-\frac{i}{\hbar }\int_{0}^{t}d\tau
E_{n}\left( \tau \right) }e^{-\int_{0}^{t}d\tau \left\langle n|\partial
_{\tau }n\right\rangle }\left\vert n(t)\right\rangle \left\langle
n(0\right\vert  , \label{s.1.1}
\end{equation}
where $\{|n(t)\rangle\}$ denotes the instantaneous eigenstate basis of a general time-dependent Hamiltonian $H_0(t)$.   
The evolution operator $U(t)$ leads an initial state $|\psi(0)\rangle = |n(0)\rangle$ into an evolved state 
$\left\vert \psi \left( t\right) \right\rangle$ given by 
\begin{equation}
\left\vert \psi \left( t\right) \right\rangle = e^{-\frac{i}{%
\hbar }\int_{0}^{t}d\tau E_{n}\left( \tau \right) }e^{-
\int_{0}^{t}d\tau \left\langle n|\partial _{\tau }n\right\rangle }\left\vert n(t) \right\rangle  , \label{xx}
\end{equation}
which mimics the adiabatic evolution of $H_0(t)$. Remarkably, such an evolution can be dictated with no adiabatic constraint by the 
 {\it superadiabatic} Hamiltonian $H_{\text{SA}}\left( t\right)$, which reads as
\begin{equation}
H_{\text{SA}}\left( t\right) =H_{0}\left( t\right) +H%
_{\text{CD}}\left( t\right) ,  \label{sfa.1.1}
\end{equation}%
where the additional term $H_{\text{CD}}\left( t\right) $ is known as the 
{\it counter-diabatic} Hamiltonian. This contribution is shown to be~\cite{Demirplak:03,Demirplak:05,Berry:09,Torrontegui:13}   
\begin{equation}
H_{\text{CD}}\left( t\right) =i\hbar \sum_{n}\left( \left\vert \partial
_{t}n\right\rangle \left\langle n\right\vert +\left\langle \partial
_{t}n|n\right\rangle \left\vert n\right\rangle \left\langle n\right\vert
\right) ,   \label{sfa.1.2}
\end{equation}%
where $\left\vert \partial _{t}n\right\rangle $ is the time derivative of $\left\vert
n(t)\right\rangle $. In particular, we have $\left\langle \partial _{t}n|n\right\rangle =0$ 
in Eq.~(\ref{sfa.1.2}) for real Hamiltonians. 
We observe that the terminology {\it superadiabaticity} has originally been introduced by Berry in Ref.~\cite{Berry:87} (see also Ref.~\cite{Berry:90}) 
as a systematic procedure of adiabatic iterations, aiming at producing successive adiabatic approximants in processes with finite slowness. 
Here, we use the term {\it superadiabatic} Hamiltonian in a different scenario, which means a Hamiltonian capable to yield a shortcut to 
adiabaticity through the presence of a counter-diabatic driving (see Ref.~ \cite{Ibanez:12,Ibanez:13} for a comparison between these two approaches). 

Note that a superadiabatic implementation of an arbitrary evolution involves  
the knowledge of the eigenstates of the adiabatic Hamiltonian $H_0\left( t\right)$. 
In some situations, this can be implemented in realizable settings. For instance, there have been 
driving protocols proposed for assisted evolutions in quantum critical phenomena~\cite{Campo:12,Campo:13,Saberi:14}. 
On the other hand, as a shortcut to accelerate QC, the application of superadiabaticity is challenging. 
Here, as we shall see, the superadiabatic implementation of gate teleportation as a primitive for universal QC 
can be promptly achieved, since we deal with the eigenspectrum of piecewise Hamiltonians, which act over a few qubits. 

\subsection{Superadiabatic teleportation of $n$-qubit states}

To derive the superadiabatic version of the teleportation of $n$-qubit states, we need to determine the counter-diabatic 
Hamiltonian $H_{\text{CD}}(s)$ associated with the Hamiltonian $H_0(s)$ as given by Eq.~(\ref{BF.1.1}). By evaluating the eigenstates of the 
blocks $H_{4\times 4}^{\pm }\left(s\right)$ in Eq.~(\ref{BF.1.5}), we get 
\begin{eqnarray}
\left\vert E_{0}^{\pm }\left( s\right) \right\rangle  &=&\left( \frac{\eta
_{i}+\chi }{\eta _{f}},\frac{\left[ \chi -\eta _{f}\right] \left[ \chi +\eta
_{i}\right] }{\eta _{i}\eta _{f}},\frac{\chi - \eta _{f}}{\eta _{i}},1\right) ,
\label{Ca-1.1.a} \\
\left\vert E_{1}^{\pm }\left( s\right) \right\rangle  &=&\left( \frac{\eta
_{i}}{\eta _{f}}-1,-\frac{\eta _{i}}{\eta _{f}},0,1\right)  , \label{Ca-1.1.b}
\\
\left\vert E_{2}^{\pm }\left( s\right) \right\rangle  &=&\left( -\frac{\eta
_{i}}{\eta _{f}},\frac{\eta _{i}}{\eta _{f}}+1,1,0\right)  , \label{Ca-1.1.c}
\\
\left\vert E_{3}^{\pm }\left( s\right) \right\rangle  &=&\left( \frac{\eta
_{i}-\chi }{\eta _{f}},\frac{\eta _{f}-\eta _{i}+\chi }{\eta _{i}-\eta _{f}+\chi },-\frac{\eta _{f}+\chi }{\eta _{i}},1\right) ,
\label{Ca-1.1.d}
\end{eqnarray}
where $\eta=\eta\left( s\right) $ and $\left\vert E_{n}^{\pm }\left( s\right)
\right\rangle $ are the non-normalized eigenstates of $H_{4\times 4}^{\pm }\left(s\right) $, 
with the function $\chi$ defined as  $\chi =\chi (s) \equiv \sqrt{\eta _{i}^{2}\left( s\right) +\eta _{f}^{2}\left( s\right) }$. 
The counter-diabatic Hamiltonian $H_{\text{CD}}(s)$ can now be found by observing that the $Z_2$ symmetries of the 
adiabatic Hamiltonian remain in the superadiabatic theory. We enunciate this result by establishing the theorem following 
(the proof is in the Appendix \ref{apendice-Teorema-simetrias}).

\begin{theorem}
Consider a time-dependent Hamiltonian $H_{0}\left( t\right) $ such that $\left[
H_{0}\left( t\right) ,\Pi _{z}\right] = 0$ and $\left[ H_{0}\left( t\right) ,\Pi _{x}%
\right] =0$, where $\Pi _{z}$ and $\Pi _{x}$ are $z$ and $x$ parity operators, respectively. Then, the superadiabatic Hamiltonian 
$H_{\text{SA}}(t)$ associated with $H_{0}\left( t\right)$ also satisfies  $\left[ H_{\text{SA}}\left( t\right) ,\Pi _{z}\right] = 0$ and 
$\left[ H_{\text{SA}}\left( t\right) ,\Pi _{x}\right] =0$. \label{teorema-simetrias}
\end{theorem}

From Theorem \ref{teorema-simetrias} we can write
\begin{equation}
H_{\text{SA}}\left( s\right) =\left[ 
\begin{array}{cc}
H_{\text{SA}}^{+}\left( s\right) & \emptyset \\ 
\emptyset & H_{\text{SA}}^{-}\left( s\right)%
\end{array}%
\right] , \label{Ca-1.11}
\end{equation}%
with $H_{\text{SA}}^{\pm }\left( s\right) \equiv H_{4 \times 4}^{\pm }\left( s\right)
+H_{\text{CD}}^{\pm }\left( s\right) $ and $H_{\text{SA}}^{+}\left( s\right)=H_{\text{SA}}^{-}\left( s\right)$. 
Since the set $\left\{ \left\vert E_{n}^{\pm
}\left( s\right) \right\rangle \right\} $ is real, we can write the counter-diabatic Hamiltonian as%
\begin{equation}
H_{\text{CD}}^{\pm }\left( s\right) =i\frac{\hbar}{\tau} \sum_{n=0}^{3}\left\vert \partial_s {E}%
_{n}^{\pm }\left( s\right) \right\rangle \left\langle E_{n}^{\pm }\left(
s\right) \right\vert  \label{Ca-1.0}
\end{equation}
 
Now, let us move on to the implementation of the superadiabatic double teleportation. To this end, we 
consider a general time-dependent Hamiltonian $H_{0}\left( s\right) $, which is split out as%
\begin{equation}
H_{0}\left( s\right) =H_{0}^{A}\left( s\right) \otimes \mathbbm{1}^{B}+\mathbbm{1}%
^{A} \otimes H_{0}^{B}\left( s\right) ,  \label{Ca-1.17.a}
\end{equation}%
where $H_{0}^{A}\left( s\right) $ and $H_{0}^{B}\left( s\right) $ are associated with piecewise superadabatic 
Hamiltonians given by $H_{\text{SA}}^{A}\left( s\right) $ and $H_{\text{SA}}^{B}\left( s\right) $,
respectively. Thus, we can write
\begin{equation}
H_{\text{SA}}\left( s\right) =H_{\text{SA}}^{A}\left( s\right) \otimes \mathbbm{1}^{B} +\mathbbm{1}^{A} \otimes H_{\text{SA}}^{B}\left( s\right) .
\label{Ca-1.17}
\end{equation}%

As a consequence, by taking the Hamiltonian of the double teleportation as given
by Eq.~(\ref{BF.2.1}), we have that the superadiabatic Hamiltonian for the double teleportation is
\begin{equation*}
H_{\text{SA}}^{D}\left( s\right) =\mathbbm{1}_{\text{even}} \otimes H_{\text{SA}}^{\text{odd}}\left( s\right)
+H_{\text{SA}}^{\text{even}}\otimes \left( s\right) \mathbbm{1}_{\text{odd}} \, ,
\end{equation*}%
where $H_{\text{SA}}^{\text{odd}}\left( s\right) $ and $H_{\text{SA}}^{\text{even}}\left( s\right) $ are
the superadiabatic Hamiltonians for each parity sector. Extension for the teleportation of 
$n$-qubit states can be achieved by adding more $Z_2$-symmetry sector, with the 
superadiabatic Hamiltonian given by 
$H_{\text{mult}}^{\text{SA}} = \sum_{k=1}^{n}  \left(\otimes_{i=1}^{k-1} \mathbbm{1}_i\right) \otimes H^{\text{SA}}_k (t) \otimes \left(\otimes_{j=k+1}^{N} \mathbbm{1}_j\right)$, 
where $H^{\text{SA}}_k (t)$ denotes the superadiabatic Hamiltonian associated with $H_k (t)$, 
with each $H_{k}\left( t\right)$ [given by Eq.~(\ref{BF.1.1})] acting on an individual sector composed by three qubits.

\subsection{Superadiabatic teleportation of $n$-qubit gates} \label{gate-teleporte}

In order to perform superadiabatic universal QC we need to show how to 
implement unitaries of one and two qubits with this model. To this end, we devise the 
the following theorem (the proof is given in the Appendix \ref{Theo2-proof}).

\begin{theorem}

Consider two time-dependent Hamiltonians $H_{0}\left( t\right) $ and $H_{0}\left(
t,G\right) $ such that $H_{0}\left(
t,G\right) =GH_{0}\left( t\right) G^{\dag }$, with $G$ denoting a unitary transformation. Then, the
superadiabatic Hamiltonian associated with $H_{0}\left( t,G\right) $ can be written as 
\begin{equation}
H_{\text{SA}}\left( t,G\right) =GH_{\text{SA}}\left( t\right) G^{\dag }  \label{Ca-1.18}
\end{equation}
where $H_{\text{SA}}\left( t\right) $ is the superadiabatic Hamiltonian of $%
H_{0}\left( t\right) $.
\label{teorema2}
\end{theorem}

Since Theorem~\ref{teorema2} holds for any unitary operator $G$ and any time-dependent 
Hamiltonian, we can use it to superadiabatically implement any unitary transformation of $n$ 
qubits. In particular, by focusing on one and two qubit gates, we can realize universal QC whose 
primitives are fast local Hamiltonians. For instance, to implement a one-qubit gate teleportation, 
the superadiabatic Hamiltonian $H_{\text{SA}}\left( t\right) $ is given by Eq.~(\ref{Ca-1.11}), while for 
the case of gate teleportation of two qubits we must consider $H_{\text{SA}}(t)$ such as given by 
Eq.~(\ref{Ca-1.17}). An important point is that, in the case of superadiabatic evolutions for rotated 
systems, the initial state is also required to be rotated (by the third party Charlie) so that the final 
state contains the teleported gate. 

\subsection{Energetic cost of superadiabatic gate teleportation}

The shortcut via a counter-diabatic Hamiltonian can yield an evolution that is faster than 
the adiabatic dynamics, but how much faster? This question has been answered for a 
general superadiabatic evolution in Ref.~\cite{alan-sarandy} through the analysis of the 
quantum speed limit (QSL) bounds~\cite{Mandelstam:45,Margolus:98,Giovannetti:03,Deffner:13} applied to superadiabatic dynamics. 
In particular, as shown in Ref.~\cite{alan-sarandy}, the total  time $\tau$ in superadiabatic evolutions 
can be arbitrarily reduced for any initial and final states as long as energy is injected in the system. More specifically, we may 
have  
$\tau \omega \rightarrow 0$, with $\omega$ denoting the energy scale of the system. 
To quantify the expense of energy in a superadiabatic evolution, we adopt the cost measure (see also Refs.~\cite{Zheng:15,Kieferova:14})
\begin{equation}
\Sigma(\tau) =\frac{1}{\tau }\int_{0}^{\tau }\left\Vert H\left( t\right)
\right\Vert dt  \, , \label{cost.1}
\end{equation}%
where $\left\Vert A\right\Vert =\sqrt{\text{Tr}\left[ A^{\dag }A\right] }$. Then, for any superadiabatic Hamiltonian $H_{\text{SA}}(t)$, 
we obtain
\begin{equation}
\Sigma(\tau) =\frac{1}{\tau }\int_{0}^{\tau }\sqrt{\sum_{m}\left[ E_{m}^{2}\left(
t\right) +\hbar ^{2}\mu _{m}\left( t\right) \right] }dt  \, , \label{cost.2}
\end{equation}%
where $\left\{ E_{m}\left(
t\right) \right\}$ is the set of energies of the adiabatic Hamiltonian $H_{0}\left( t\right) $ and%
\begin{equation}
\mu _{m}\left( t\right) =\left\langle \partial_t{m}\left(t\right) |\partial_t{m}\left( t\right) \right\rangle-\left\vert \left\langle m\left(t\right) | \partial_t{m}\left( t\right) \right\rangle \right\vert ^{2} \, .
\label{cost.4}
\end{equation}%
Equation (\ref{cost.2}) shows an increase in the energetic cost to superadiabatic evolutions compared 
to their adiabatic counterparts. Let us now evaluate the energetic cost to implement universal QC 
via teleportation. To this end, we calculate first the cost of single and double state teleportation 
and then extend the analysis for the cost of the implementation of quantum gates. 
By parametrizing the evolution in terms of the normalized time $s=t/\tau $, the energetic cost $\Sigma_{single}$ 
for the teleportation of a single qubit reads as
\begin{equation}
\Sigma_{single}=\int_{0}^{1}\sqrt{\sum_{m}\left[ E_{m}^{2}\left( s\right)
+\hbar ^{2} \frac{\mu _{m}\left( s\right) }{\tau ^{2}}\right] }ds \, ,
\label{cost.5}
\end{equation}%
where $\mu _{m}\left( s\right) = \left\langle
\partial_s E_{m}\left( s\right) |\partial_s E_{m}\left( s\right) \right\rangle$, which is a consequence of the fact that the set of 
eigenvalues of $H_{0}\left( s\right)$ is real. To illustrate the dependence of the energetic cost on the evolution path 
adopted, we will choose three interpolations: (i) linear interpolation, with $\eta _{i}\left( s\right)
=1-s$ and $\eta _{f}\left( s\right) =s$; (ii) trigonometric interpolation, with $\eta _{i}\left( s\right) =\cos \left( \pi s/2\right)$ and  
$\eta_{f}\left( s\right) =\sin \left( \pi s/2\right) $; and (iii) exponential interpolation, with $\eta _{i}\left( s\right) =\left( e^{1-s}-1\right)
/\left( e-1\right) $ and $\eta _{f}\left( s\right) =\left( e^{s}-1\right)
/\left( e-1\right) $.
\begin{figure}[!htb]
\centering
\includegraphics[scale=0.3]{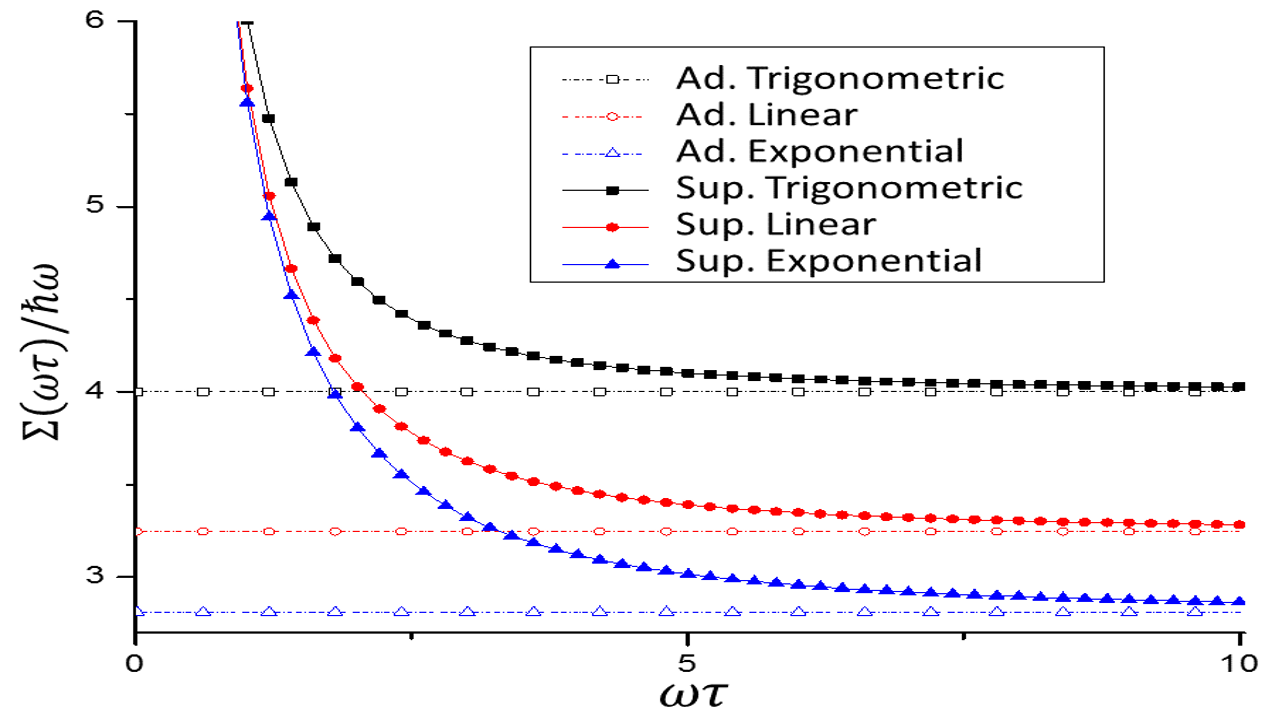}
\caption{(Color online) Energetic cost as a function of $\tau \omega $ for both adiabatic and superadiabatic dynamics of single qubit teleportation. 
Notice that the superadiabatic cost recovers the cost of its adiabatic counterpart in the limit $\tau \omega \rightarrow \infty$.}
\label{graph1}
\end{figure}

Then, we numerically evaluate the energetic cost as a function of $ \omega \tau$ of by applying Eq.~(\ref{cost.5}) to each interpolation, 
which is plotted in Fig.~\ref{graph1}. In this plot, we explicitly show that the superadiabatic evolution recovers the cost of its adiabatic 
counterpart at the limit of infinite $\omega \tau$. Notice also that the usual linear interpolation is {\it not} the less costly option of 
interpolation. Moreover, the plot is in agreement with the energy-time complementarity relationship, with the faster evolutions costing more energy than 
slower dynamics.
The energetic cost to implement the superadiabatic teleportation of an unknown
$n$-qubit state can be provided in terms of the cost to implement the
single teleportation as (see Appendix \ref{prova-custo})%
\begin{equation}
\Sigma _{n}=g_{n} \, \Sigma _{single}  , \label{costN}
\end{equation}%
where we define the function $g_{n}=\sqrt{2^{3\left( n-1\right) }n}$.
Moreover, the cost to implement {\it gate} teleportation of $n$ qubits via superadiabatic
evolution is also given by Eq.~(\ref{costN}) due to the invariance of the
Hilbert-Schmidt norm by unitary rotations. Note that the factor $g_{n}$ exponentially increases with $n$. 
In any case, this is not a problem to perform universal QC with one and two qubits. In that case, we have 
$g_{2}=$ $4$ and $g_{3}=8\sqrt{3}$, respectively.

\section{Conclusion} \label{Conc}

We introduced a general shortcut to the adiabatic gate teleportation model of quantum computation. 
Moreover, the model has been generalized to include the teleportation of 
an arbitrary $n$-qubit unitary gate. In particular, we have shown through 
Theorem~\ref{teorema2} that the superadiabatic Hamiltonian for the 
teleportation of an $n$-qubit state can be directly used to implement the 
teleportation of an $n$-qubit gate $U$ through a simple $U$ rotation over the 
original superadiabatic Hamiltonian. 
As a main result of the work, we have shown that it is possible to devise 
fast local Hamiltonians to perform teleportation of one and two qubits as a primitive of universal QC. 
To analyze the energetic cost of the superadiabatic evolution, 
we considered the time-energy complementary relationship. In this context, it has been 
shown that the superadiabatic implementation is always more costly than its 
adiabatic counterpart, reducing to it in the limit of a long evolution time. 

Implications of the superadiabatic approach applied to gate teleportation in 
a decohering environment is a further challenge of interest. In open systems, 
there is a competition between the adiabatic time scales, which require a long 
evolution, and the decoherence characteristic times, which require fast evolution. 
In this scenario, the superadiabatic implementation may provide a direction 
to obtain an optimal running time for the quantum algorithm while keeping an 
inherent protection against decoherence. A basis for such analysis may be 
provided by the generalization of the superadiabatic theory for the context of 
open systems (see, e.g., Refs.~\cite{Jing:13,Vacanti:14,Jing:15,Song:15}). 
The robustness of superadiabatic gate teleportation as well as experimental 
proposals are left for future research.      

\begin{acknowledgments}
We thank Steven Flammia for useful discussions. 
M. S. S. thanks Daniel Lidar for his hospitality at the University of Southern California.  
We acknowledge financial support from the Brazilian agencies CNPq, CAPES, FAPERJ,  
and the Brazilian National Institute of Science and Technology for Quantum Information (INCT-IQ).
\end{acknowledgments}

\appendix

\section{Proof of the Theorem \ref{teorema-simetrias}} \label{apendice-Teorema-simetrias}

The proof of Theorem \ref{teorema-simetrias} can be obtained as follows. 
If a time-dependent Hamiltonian $H_{0}\left( t\right)$ satisfies the commutation 
relation $\left[ H_{0}\left( t\right) ,\Pi _{z}\right] =0$, then we can write 
$\left[ H_{\text{SA}}\left( t\right) ,\Pi _{z}\right] =\left[ H_{\text{CD}}\left(t\right) ,\Pi _{z}\right] $. 
As $\Pi _{z}$ and $H_{0}\left( t\right) $ have a common basis of eigenstates, an 
eigenstate $\left\vert n\left( t\right) \right\rangle $ of $H_{0}\left( t\right) $ has a 
definite $\Pi _{z}$ parity so that we can write $\Pi _{z}\left\vert n\left( t\right) \right\rangle =\left( -1\right)
^{n}\left\vert n\left( t\right) \right\rangle $ (by encoding the parity into the label $n$). 
By using $\Pi_{z}\left\vert \partial_t{n}\left( t\right) \right\rangle =\left( -1\right)
^{n}\left\vert \partial_t{n}\left( t\right) \right\rangle $ it follows that $H_{\text{CD}}\left( t\right) \Pi _{z}=\Pi _{z}H_{\text{CD}}\left( t\right) $, 
thus implying $\left[ H_{\text{SA}}\left( t\right) ,\Pi _{z}\right] =0$.
To complete the demonstration, from the hypothesis that  
$\left[ H_{0}\left( t\right) ,\Pi _{x}\right] =0$ is satisfied, we write $\left[ H_{\text{SA}}\left( t\right) ,\Pi _{x}\right] =\left[
H_{\text{CD}}\left( t\right) ,\Pi _{x}\right] $. Then, let us denote a matrix element of $\left[ H_{\text{CD}}\left( t\right) ,\Pi _{x}%
\right] $ in the basis of eigenstates of $H_{0}\left( t\right) $ as 
\begin{equation}
\left[ H_{\text{CD}}\left( t\right) ,\Pi _{x}\right] _{kl} = 
\left\langle k\left( t\right) |\left[ H_{\text{CD}}\left( t\right) ,\Pi _{x}\right] |l\left( t\right)
\right\rangle . 
\end{equation}
We now use that $\Pi _{x}\left\vert n\left( t\right) \right\rangle =\left\vert {n}^\prime\left( t\right) \right\rangle$, 
where $\left\vert n\left( t\right)\right\rangle $  and $\left\vert n^{\prime }\left( t\right) \right\rangle $ are eigenstates of 
the parity operator $\Pi _{z}$, with opposite eigenvalues. Moreover, $\Pi_{x}\left\vert \partial_t{n}\left( t\right) \right\rangle =\left\vert \partial_t{n}%
^{\prime }\left( t\right) \right\rangle $. Then
\begin{eqnarray}
\left(H_{\text{CD}}\left( t\right) \Pi _{x}\right) _{kl} &=& i\hbar \left[ \left\langle k \left( t\right) | \partial_t{l}^\prime\left( t\right)
\right\rangle + \left\langle \partial_t{l}^\prime\left( t\right) |l\left( t\right)
\right\rangle \left\langle k\left( t\right) |l^{\prime }\left(
t\right) \right\rangle \right] \nonumber \\
&=& \left( \Pi _{x} H_{\text{CD}}\left( t\right) \right) _{kl}. 
\end{eqnarray}
Thus $\left[ H_{\text{CD}}\left( t\right) ,\Pi _{x}\right] _{kl}=0$ $\forall $ $%
\left( k,l\right) $. This proves Theorem \ref{teorema-simetrias} .

\section{Proof of the Theorem \ref{teorema2}}
\label{Theo2-proof}

In order to prove Theorem \ref{teorema2}, consider two Hamiltonians $H\left( t\right) 
$ and $H\left( t,G\right) $ such that $H\left( t,G\right) =GH\left( t\right)
G^{\dag }$, with $GG^{\dag }=\1$. The set of
eigenvectors $\left\vert n\left( s\right) ,G\right\rangle $ of the
Hamiltonian $H\left( t,G\right) $ can be determined from the set of
eigenvectors $\left\vert n\left( s\right) \right\rangle $ of adiabatic
Hamiltonian $H\left( s\right) $ as follows 
\begin{equation}
\left\vert n\left( t\right) ,G\right\rangle =G\left\vert n\left( t\right)
\right\rangle  .  \label{CDn}
\end{equation}
Thus, the counter-diabatic Hamiltonian associated with $H\left( s,G\right) $ is
given by%
\begin{equation}
H_{\text{CD}}\left( s,G\right) =\frac{i\hbar }{\tau }\sum_{n}\left\vert \partial
_{s}n,G\right\rangle \left\langle n,G\right\vert +\left\langle \partial
_{s}n,G|n,G\right\rangle \left\vert n,G\right\rangle \left\langle
n,G\right\vert   . \label{sfa.1.22}
\end{equation}%
Then, by using Eq.~(\ref{CDn}), we can show that%
\begin{equation}
H_{\text{CD}}\left( s,G\right) =G\left[ \frac{i\hbar }{\tau }\sum_{n}\left\vert
\partial _{s}n\right\rangle \left\langle n\right\vert +\left\langle \partial
_{s}n|n\right\rangle \left\vert n\right\rangle \left\langle n\right\vert %
\right] G^{\dag } ,
\end{equation}%
where we have used that $\left\vert \partial _{s}n,G\right\rangle
=G\left\vert \partial _{s}n\right\rangle $ and $GG^{\dag }=\1$. Hence, we
can write%
\begin{equation}
H_{\text{CD}}\left( s,G\right) =GH_{\text{CD}}\left( s\right) G^{\dag } .
\label{theo-proof}
\end{equation}
Eq.~(\ref{theo-proof}) implies that $H_{\text{SA}}\left( t,G\right) =GH_{\text{SA}}\left( t\right) G^{\dag }$. 
This proves Theorem~\ref{teorema2}.

\vspace{0.3cm}

\section{Proof of Eq.~(\ref{costN})} \label{prova-custo}

In order to demonstrate Eq.~(\ref{costN}), let us write the
adiabatic Hamiltonian that is used to perform the $n$-qubit state teleportation as
\begin{equation}
H_{\text{SA}}\left( s\right) =\sum\limits_{k=1}^{n}\mathcal{H}_{k}^{\text{SA}}\left(
s\right) 
\end{equation}
where $\mathcal{H}_{k}^{\text{SA}}\left( s\right) =\left( \otimes
_{l=1}^{k-1}\1_{l}\right) \otimes H_{k}^{\text{SA}}\left( s\right) \otimes \left( \otimes
_{l=k+1}^{n}\1_{l}\right) $, with $\mathcal{H}_{k}^{\text{SA}}\left( s\right) $ being a 
three-qubit Hamiltonian for each independent sector, as displayed in Fig.~\ref{figure3} . 
Then, the energetic cost for the $n$-qubit superadiabatic teleportation reads as
\begin{equation}
\Sigma _{n}=\int_{0}^{1} ds \sqrt{\text{Tr}\left[ H_{\text{SA}}^{2}\left( s\right) \right] } ,
\label{CostN}
\end{equation}%
where we can write%
\begin{equation}
H_{\text{SA}}^{2}\left( s\right) =\sum\limits_{k=1}^{n}\left[ \mathcal{H}%
_{k}^{\text{SA}}\left( s\right) \right] ^{2}+\sum\limits_{m\neq k} \left(\sum_{k} \mathcal{H}_{k}^{\text{SA}}\left( s\right) \mathcal{H}%
_{m}^{\text{SA}}\left( s\right) \right) . 
\end{equation}
Now, we use that, for $k \ne m$, we get 
\begin{equation}
\text{Tr}\left[ \mathcal{H}_{k}^{\text{SA}}\left( s\right) \mathcal{H}%
_{m}^{\text{SA}}\left( s\right) \right] = 
\left(\text{Tr}\left[\mathbbm{1}\right]\right)^{n-2} \, \text{Tr}\left[ {H}_{k}^{\text{SA}}\left(
s\right) \right] \, \text{Tr}\left[ {H}_{m}^{\text{SA}}\left( s\right) \right] . 
\end{equation}
Then, we write $\text{Tr}\left[ {H}_{j}^{\text{SA}}\left( s\right) \right] = \text{Tr}\left[ {H}^{(0)}_{j}\left( s\right)+ {H}^{\text{CD}}_{j}\left( s\right)  \right]$, 
where ${H}^{(0)}_{j}\left( s\right)$ is the original (adiabatic) Hamiltonian at sector $j$ and ${H}^{\text{CD}}_{j}\left( s\right)$ its corresponding 
counter-diabatic Hamiltonian. By explicitly computing the trace in 
the eigenstate basis of  ${H}^{(0)}_j(s)$ and 
by using Eqs.~(\ref{BF.1.9a})-(\ref{BF.1.9c}) and (\ref{sfa.1.2}), we obtain that $\text{Tr}\left[ {H}_{j}^{\text{SA}}\left( s\right) \right] =0$ ($\forall j \in \{1,\cdots,n\}$), 
which implies
\begin{equation}
\text{Tr}\left[ \mathcal{H}_{k}^{\text{SA}}\left( s\right) \mathcal{H}%
_{m}^{\text{SA}}\left( s\right) \right] = 
0 \,\,\,\,\, ( k \ne m) \,.
\end{equation}
Thus, the energetic cost for the $n$-qubit state teleportation reads as
\begin{eqnarray}
\text{Tr}\left[ H_{\text{SA}}^{2}\left( s\right) \right] &=& \sum\limits_{k=1}^{n}\text{Tr}\left\{ \left[ \mathcal{H}_{k}^{\text{SA}}\left( s\right) \right] ^{2}\right\} \nonumber \\
&=& \left(\text{Tr}\left[\mathbbm{1}\right]\right)^{n-1} \, \sum\limits_{k=1}^{n} \text{Tr}\left\{ \left[ H_{k}^{\text{SA}}\left( s\right) \right] ^{2} \right\}  \nonumber \\
&=& 2^{3\left( n-1\right) }\, n \, \text{Tr}\left\{ \left[ H_{k}^{\text{SA}}\left( s\right) \right]^{2}\right\}  \,\,\,  (\forall k )   \, . \label{Cfinal}
\end{eqnarray}
Hence, Eq.~(\ref{Cfinal}) into Eq.~(\ref{CostN}) yields
\begin{equation}
\Sigma _{n}=\sqrt{2^{3\left( n-1\right) }n}\Sigma_{single} \, ,
\end{equation}%
which proves the validity of Eq.~(\ref{costN}).


\end{document}